\documentclass[conference]{IEEEtran}
\usepackage{epsfig}
\usepackage{graphicx}
\usepackage{listings}
\usepackage{url}
\usepackage{cite}
\usepackage{alltt}
\usepackage{balance}
\usepackage{wrapfig}
\usepackage{multirow}

\usepackage{times}
\usepackage{cite}
\usepackage{graphicx}
\usepackage{epsfig}
\usepackage{hyperref}
\usepackage{xspace}
\usepackage{latexsym}
\usepackage[usenames]{color}
\usepackage{multirow}
\usepackage{listings}
\usepackage{textcomp}

\def\denseitems{
  \itemsep1pt plus1pt minus1pt
  \parsep0pt plus0pt
  \parskip0pt\topsep0pt}

\makeatletter
\newcommand\tabcaption{\def\@captype{table}\caption}
\makeatother

\hypersetup{
colorlinks=false,
bookmarksnumbered=true,
}

\newcommand{\ie}{\textit{i.e.,} }
\newcommand{\eg}{\textit{e.g.,} }

\newcommand\opt[1]{}
\newcommand\find[1]{}

\def\denseitems{
  \itemsep1pt plus1pt minus1pt
  \parsep0pt plus0pt
  \parskip0pt
  \topsep0pt
}
\newcommand{\ls}[1]
   {\dimen0=\fontdimen6\the\font 
    \lineskip=#1\dimen0
    \advance\lineskip.5\fontdimen5\the\font
    \advance\lineskip-\dimen0
    \lineskiplimit=.9\lineskip
    \baselineskip=\lineskip
    \advance\baselineskip\dimen0
    \normallineskip\lineskip
    \normallineskiplimit\lineskiplimit
    \normalbaselineskip\baselineskip
    \ignorespaces
   }

\newenvironment{smalldescription}{
   \setlength{\topsep}{0pt}
   \setlength{\partopsep}{0pt}
   \setlength{\parskip}{0pt}
   \begin{description}
   \setlength{\leftmargin}{.2in}
   \setlength{\parsep}{0pt}
   \setlength{\parskip}{0pt}
   \setlength{\itemsep}{0pt}}{\end{description}}

\newcounter{observation}
\usepackage[linesnumbered,boxed,ruled,nofillcomment]{algorithm2e} 

\usepackage{amsmath,amsthm}

\begin{document}

\abovedisplayskip=3pt
\belowdisplayskip=3pt
\abovedisplayshortskip=0pt
\belowdisplayshortskip=0pt


\title{Improving Efficiency and Scalability of\\Formula-based Debugging}

 \author{\IEEEauthorblockN{Wei Jin and Alessandro Orso}
 \IEEEauthorblockA{
 Georgia Institute of Technology\\
 \{weijin\textbar{}orso\}@gatech.edu}
 }

\maketitle



\begin{abstract}
  Formula-based debugging techniques are becoming increasingly
  popular, as they provide a principled way to identify potentially
  faulty statements together with information that can help fix such
  statements. Although effective, these approaches are computationally
  expensive, which limits their practical applicability. Moreover,
  they tend to focus on failing test cases alone, thus ignoring the
  wealth of information provided by passing tests. To mitigate these
  issues, we propose two techniques: on-demand formula computation
  (OFC) and clause weighting (CW). OFC improves the overall efficiency
  of formula-based debugging by exploring all and only the parts of a
  program that are relevant to a failure. CW improves the accuracy of
  formula-based debugging by leveraging statistical fault-localization
  information that accounts for passing tests.  Our empirical results
  show that both techniques are effective and can improve the state of
  the art in formula-based debugging.
\end{abstract}


\section{Introduction}
\label{sec:intro}

Because debugging is expensive and time consuming, there has been a
great deal of research on automated techniques for supporting various
debugging tasks (\eg \cite{Zeller2002TSE, jones2002, Liblit2003,
  liblit05, Abreu2008, Santelices2009, Liu2005, Artzi2010, Zhang2009,
  Baah2011}). Recently, in particular, there has been a considerable
interest in techniques that can perform fault localization in a more
principled way (\eg \cite{Jose2011, Torlak2010, Ermis2012,
  Christ13}). These techniques, collectively called
\textit{formula-based debugging}, model faulty programs and failing
executions as formulas and perform fault localization by manipulating
and solving these formulas. As a result, they can provide developers
with the possible location of the fault, together with a mathematical
explanation of the failure (\eg the fact that an expression should
have produced a different value or that a different branch should have
been taken at a conditional statement).

BugAssist~\cite{Jose2011} is a technique of particular interest in
this arena. Given a faulty program, a failing input, and a
corresponding (violated) assertion, BugAssist performs fault
localization by constructing an unsatisfiable Boolean formula that
encodes (1) the input values, (2) the semantics of (a bounded version
of) the faulty program, and (3) the assertion. It then uses a pMAX-SAT
solver to find maximal sets of clauses in this formula that can be
satisfied together and outputs the complement sets of clauses (CoMSS)
as potential causes of the error. Intuitively, each set of clauses in
CoMSS indicates a corresponding set of statements that, if suitably
modified (\eg replacing the statements with angelic
values~\cite{chandra2011}), would make the program behave correctly
for the considered input.

Although effective, BugAssist is extremely computationally expensive,
as it builds a formula for (a bounded unrolling of) \textit{all
  possible paths} in a program. This can lead to formulas with
millions of terms~\cite{Jose2011} and scalability issues even for
small programs. Moreover, BugAssist, like most formula-based debugging
approaches, does not take into account passing test cases, thus
missing two important opportunities.  First, passing executions can
help identify statements, and thus parts of the formulas, that are
less likely to be related to the fault, which can help optimizing the
search for a solution to such formulas. Second, passing executions can
help filtering out locations that may be potential fixes for the failing
executions considered but could break previously passing test cases if
modified~\cite{chandra2011}.

In this paper, we propose two possible ways of addressing these issues
and improving formula-based debugging approaches: \textit{on-demand
  formula computation (OFC)} and \textit{clause weighting (CW)}.
OFC is a novel on-demand algorithm that can dramatically reduce the
number of paths encoded in a formula, and thus the overall complexity
of such formula and the cost of computing a pMAX-SAT solution for it.
Intuitively, our algorithm (1) builds a formula for the path in the
original failing trace, (2) analyzes the formula to identify
additional relevant paths to consider, (3) expands the formula by
encoding these additional paths, (4) repeats (2) and (3) until no more
relevant paths can be identified, at which point it (5) reports the
computed solution.
CW accounts for the information provided by passing test cases by
assign weights to the different clauses in an encoded formula based on
the suspiciousness values computed by a statistical fault localization
technique. Doing so has the potential to improve the accuracy of the
results by helping the solver compute CoMSSs that are more likely to
correspond to faulty statements.  (The guidance provided to the solver
can also unintentionally improve the efficiency of the approach, as we
show in Section~\ref{sec:rq1-weighted}.)
To assess the effectiveness of OFC and CW, we selected BugAssist as a
baseline and considered four different formula-based debugging
techniques: the original BugAssist, BugAssist+CW, OFC, and OFC+CW.  We
implemented all four techniques in a tool that works on C programs and
used the tool to perform an empirical study.  In the study, we first
applied the four techniques to $52$ versions of two small programs to
assess several tradeoffs involved in the use of CW and OFC and compare
with related work. Our results are encouraging, as they show that CW
and OFC can improve the performance of BugAssist in several respects.
First, the use of CW resulted in more accurate results---in terms of
position of the actual fault in the ranked list of statements reported
to developers---in the majority of the cases considered. Second, CW
and OFC were able to reduce the computational cost of BugAssist by
27\% and 75\% on average, respectively, with maximum speedups of over
70X for OFC. Most importantly, our results show that these
improvements, and especially OFC, allow formula-based debugging to
handle faults that go beyond the capability of an all-paths analysis
such as the one performed by BugAssist. To further demonstrate the
practicality of CW and OFC, we also performed a case study on a
real-world bug in Redis, a popular open source project. Overall, our
results show that CW and OFC are promising steps towards more
practically applicable formula-based debugging techniques and motivate
further research in this direction.

The main contributions of this paper are:

\begin{itemize}\denseitems

\item The definition of clause weighting and on-demand formula
  computation, two approaches for improving the accuracy and
  efficiency of formula-based debugging.

\item A prototype implementation of our technique that is available
  for download, together with our experimental infrastructure and
  benchmark programs (see {\small
    \url{http://www.cc.gatech.edu/~orso/software/odin/}}).

\item Initial empirical evidence that CW and OFC are as effective, more 
  efficient, and potentially more practically applicable than existing 
  approaches. 


\end{itemize}




\section{Background}
\label{sec:bg}

\paragraph*{SSA Form}
\label{sec:bg:ssa}

Given a program $P$, the static single assignment (SSA) form of $P$ is
a program semantically equivalent to $P$ in which each variable is
assigned exactly once~\cite{Cytron1991}. Because multiple definitions
can reach a join point, for each conditional statement $cs$, the SSA
form contains one $\phi$ function \textit{phi} for each definition $d$
in the original program that is control dependent on $cs$ and can
reach $cs$'s join point.  \textit{phi} is located at the join point
and selects the correct definition to use at that point depending on
which branch of $cs$ was executed.  We refer to conditional statement
$cs$ as \textit{phi}'s \textit{conditional}.
%

\paragraph*{Statistical Fault Localization}

Spectrum-based statistical fault localization techniques compute the
correlation between a program entity and an observed failure based on
how the entity was exercised in passing and failing executions (\eg
\cite{jones2002, liblit05, ochiai}). They use this correlation as an
approximation of the likelihood of a program entity to be faulty. We
leverage the Ochiai fault localization approach, which has shown to be
quite effective in empirical comparisons~\cite{ochiai}.

\paragraph*{MAX-SAT, pMAX-SAT, and wpMAX-SAT Problems}

MAX-SAT is the problem of determining the maximum number of clauses of
a given unsatisfiable Boolean formula that can be satisfied by some
assignment~\cite{Bailey2005}. An extension of MAX-SAT is pMAX-SAT, in
which clauses are marked as either hard (\ie clauses that cannot be
dropped) or soft (\ie clauses that can be dropped). wpMAX-SAT extends
pMAX-SAT by assigning weights to soft clauses, such that clauses with
higher weights are less likely to be dropped.
A solution to a wpMAX-SAT problem is a maximal satisfiable subset of
clauses (MSS) with maximum weight in which all hard clauses are
satisfied. The complement of MSS is called CoMSS. MSS is defined as a
maximal set of clauses, in the sense that adding any of the other
clauses in CoMSS would make the set unsatisfiable. The maximal
property of MSS and the minimal property of CoMSS essentially imply
that clauses in CoMSS are responsible for making the formula
unsatisfiable. There may be several different maximal satisfiable
subsets and complementary sets for a given MAX-SAT problem, and each
of these sets can contain multiple clauses.

\section{Improving Formula-based Debugging}
\label{sec:impr-form-based}

As we discussed in the Introduction, the goal of this work is to
investigate ways to mitigate some of the limitations of existing
formula-based debugging approaches. To this end, we propose two
approaches: clause weighting and on-demand formula computation. We
discuss them in detail using BugAssist~\cite{Jose2011} as a
representative of state-of-the-art formula-based debugging techniques
and our baseline.

\subsection{Clause Weighting (CW)}
\label{sec:weight-computation}

CW consists of using the information from passing executions to inform
a wpMAX-SAT solver. More precisely, CW leverages the suspiciousness
values computed by a statistical fault localization technique and
assigns to each program entity $en$, and thus to the corresponding
clause in the program formula, a weight inversely proportional to its
suspiciousness {\scriptsize $susp(en)$}: {\scriptsize $weight(en) =
  1/susp(en)$}. If the suspiciousness value of an entity is zero,
which means that the entity is only executed by passing tests, CW
assigns to it the largest possible weight. By assigning different
weights to different clauses, CW transforms the original pMAX-SAT
problem in BugAssit into a wpMAX-SAT problem. The rationale for CW is
that, by the definition of wpMAX-SAT, clauses with higher weights are
more likely to be included in an MSS (\ie less likely to be identified
as causes of the faulty behavior), while clauses with lower weights
are less likely to be included in an MSS (\ie more likely to be
included in a CoMSS and thus be identified as causes of the faulty
behavior).


\begin{figure*}[t]
\centering
\includegraphics[width=1.5\columnwidth,keepaspectratio]{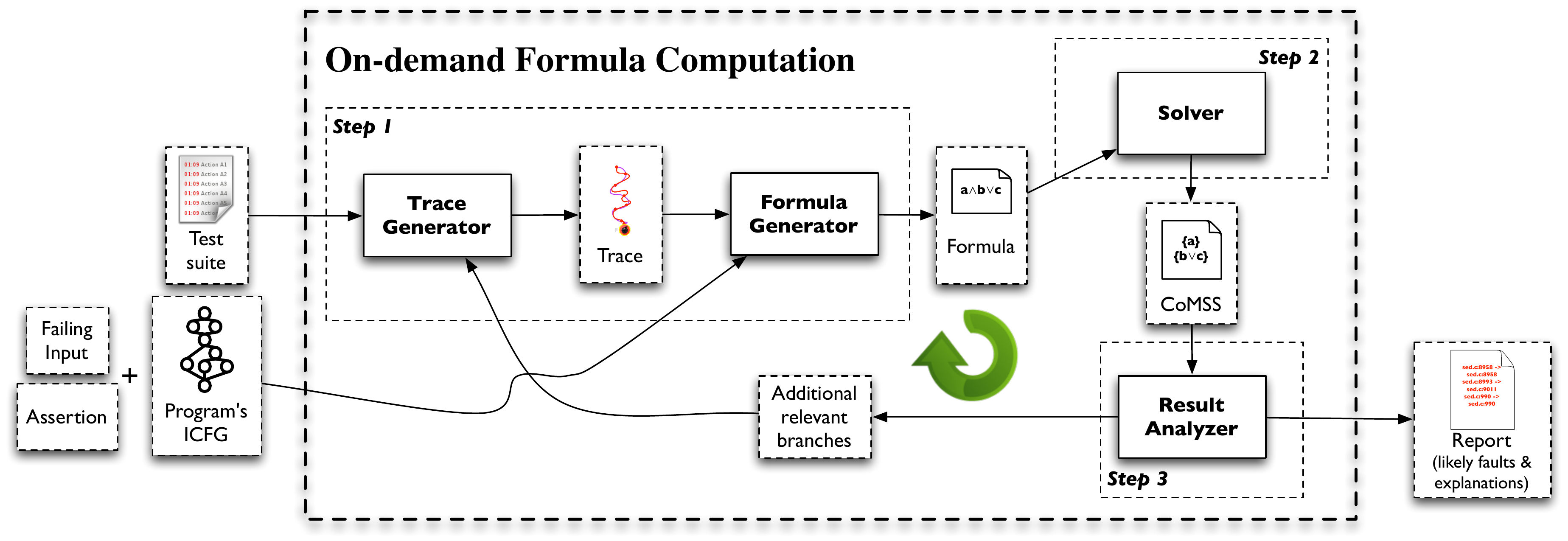}
\caption{Overview of on-demand formula computation.}
\label{fig:overall}
\end{figure*}

Formula-based debugging techniques such as BugAssist consider all
possible pMAX-SAT solutions equally and simply report them.
Conversely, by leveraging the heuristics in statistical fault
localization, CW is more likely to rank the set of clauses
corresponding to the fault at the top of the list of solutions, thus
reducing developers' debugging effort. This potential advantage,
however, comes at a cost. Solving wpMAX-SAT problems can be
computationally more expensive than solving a pMAX-SAT problem, which
can outweigh CW's benefits. To understand this tradeoff, in our
empirical evaluation we assess how CW affects the accuracy and
efficiency of formula-based debugging (see
Section~\ref{sec:rq1-weighted}).

\subsection{On-demand Formula Computation (OFC)}
\label{sec:algo}

OFC is our second, and more substantial, improvement over traditional
formula-based debugging techniques. Figure~\ref{fig:overall} shows an
overall view of OFC and its workflow. The inputs to the algorithm are
a faulty program, represented as an Inter-procedural Control Flow
Graph (ICFG), and a test suite that contains a set of passing tests
and one failing test. (We discuss how OFC could leverage the presence of multiple
failing inputs in Section~\ref{sec:furth-optim}.) As it is common practice for debugging
techniques, we assume that a failure can be expressed as the violation
of an assertion in the program. Given these inputs, OFC produces as
output a set of clauses and their corresponding program entities (\ie
branches and statements). These are entities that, if suitably
modified, would make the failing execution pass. The expressions in
the reported clauses provide developers with additional information on
the failure, and can be considered a ``mathematical explanation'' of
the failure.

As Figure~\ref{fig:overall} shows, OFC consists of three main steps.
The key idea behind OFC is to reason about the failure (and the
program) incrementally, by starting with the entities traversed in a
single failing trace, computing CoMSS solutions for the partial
program exercised by the trace, and then expanding the portion of the
program considered in the analysis when such solutions indicate that
additional control-flow paths should be taken into consideration to
``explain'' the failure. Specifically, in its first step
(Section~\ref{sec:trace-gener-form}), OFC generates a new trace (the
original failing trace, in the first iteration) and suitably updates
the trace formula, a formula that encodes the semantics of the traces
generated so far. OFC's second step (Section~\ref{sec:solver})
computes the CoMSSs of the (unsatisfiable) formula built in the
previous step. Finally, in OFC's third step, the algorithm checks
whether there is any additional relevant branch to consider in the
program (Section~\ref{sec:result-analyzer}). If so, OFC returns to
Step~1. Otherwise, it computes all possible CoMSSs of the final
formula to report to developers the set of relevant clauses and their
corresponding program entities.

Algorithm~\ref{alg:odin_main} shows the main algorithm, which takes as
inputs the ICFG of the faulty program and the program's test suite and
performs the three steps we just described. We discuss each step in
detail in the rest of this section.


\subsubsection{Trace Generator and Formula Generator}
\label{sec:trace-gener-form}

After an initialization phase, OFC iterates Steps~1,~2, and~3.  Step~1
performs two tasks: trace generation and formula generation.

\paragraph*{Trace Generator}

In its first part, Step~1 invokes the \textit{Trace Generator}
(Algorithm~\ref{alg:tracegenerator}). In the first iteration of the
algorithm, \textit{Trace Generator} generates the trace corresponding
to the failing input. In subsequent iterations, it generates a trace
that covers the new program entities identified as relevant by Step~3
(see Section~\ref{sec:result-analyzer}), so as to augment the scope of
the analysis. The inputs to \textit{TraceGenerator} are the failing
input, the map that associates each branch covered so far with the
trace in which it was first covered, and the new relevant branch for
which a trace must be generated (by flipping it).

If \textit{flip\_br} is null, which only happens in the first
iteration of the algorithm, \textit{TraceGenerator} generates a trace
by simply providing the failing input to the program and collecting
its execution trace (line~\ref{alg:execute1}). Otherwise, for
subsequent iterations, \textit{TraceGenerator} retrieves
\textit{old\_trace} (line~\ref{alg:execute2}), the trace that first
reached branch \textit{flip\_br} and generates a new trace,
\textit{new\_trace} (line~\ref{alg:execute2}). To generate the trace,
the algorithm provides the failing input to the program, forces the
program to follow \textit{old\_trace} up to \textit{flip\_br}, and
flips \textit{flip\_br} so that the program follows its alternative
branch (using execution hijacking~\cite{tsankov2011execution}). The
algorithm also updates map \textit{visited\_branches} by adding to it
an entry for every branch newly covered by \textit{new\_trace},
including \textit{flip\_br}'s alternative branch
(lines~\ref{alg:insertbranch1}--\ref{alg:insertbranch2}).

\paragraph*{Formula Generator}
\label{sec:alg:encode}

After generating a trace, OFC invokes \textit{FormulaGenerator}
(Algorithm~\ref{alg:encodefunction}), which constructs a new formula
\textit{TF}, either from scratch (in the first iteration) or by
expanding the current formula based on the program entities in
\textit{new\_trace} (in subsequent iterations).

The inputs to \textit{FormulaGenerator} are the ICFG of the faulty
program, the current trace formula, the portion of the program
currently considered (and encoded in the current trace formula), the
trace newly generated by \textit{TraceGenerator}, and a map from
clauses to statements that originated them.

In its main loop, \textit{FormulaGenerator} processes each statement
\textit{st} in the new trace, \textit{new\_trace}, one at a time. If
\textit{st} is not yet part of \textit{SP}, the portion of the program
currently considered, the algorithm (1) adds \textit{st} to
\textit{SP}, (2) encodes its semantics in a new Boolean clause
\textit{clause$_{st}$}, (3) conjoins \textit{clause$_{st}$} and
\textit{TF}, and (4) updates map \textit{clause\_origin} by mapping
\textit{clause$_{st}$} to \textit{st}.

Similar to other symbolic analyses (\eg~\cite{ckl2004, Jose2011,
  Torlak2010}), OFC operates on an SSA form of the faulty program (see
Section~\ref{sec:bg:ssa}). The formula generator models three types of
statements in the program (and its trace): conditional statements (\eg
line~$1$ in Figure~\ref{fig:example1}), definitions that involve a
$\phi$ function (\eg line~$phi1$ in Figure~\ref{fig:example1} (right))
and definitions that do not involve a $\phi$ function. Intuitively,
whereas the last type of statements represent traditional data-flow
information about uses and definitions, the other two types encode
control-flow information about branch conditions and $\phi$ function
selection conditions. To perform a correct semantic encoding, when
deriving \textit{clause$_{st}$} from \textit{st},
\textit{FormulaGenerator} must treats these three types of statements
differently.

If \textit{st} is a conditional statement with predicate
\textit{predicate$_{st}$}, the algorithm retrieves such predicate from
\textit{st} (line~\ref{alg:getpred}) and encodes \textit{st} as
\textit{(guard$_{st}$=predicate$_{st}$)}, where \textit{guard$_{st}$}
is a Boolean variable that represents \textit{st}'s condition
(line~\ref{alg:predassignment}).

If \textit{st} involves a $\phi$ function $phi$, the algorithm
generates a clause $(\mathit{guard_{cs}} \wedge (st_{LHS}=st_{RHS,t}))
\vee (\neg \mathit{guard_{cs}} \wedge (st_{LHS}=st_{RHS,f}))$, where
(1) \textit{cs} is \textit{phi}'s conditional and, similar to above,
$\mathit{guard_{cs}}$ represents \textit{cs}'s condition, (2)
$st_{LHS}$ is the variable being defined at \textit{st}, and (3)
$st_{RHS,t}$ and $st_{RHS,f}$ are the definitions selected by
\textit{phi} along \textit{cs}'s \textit{true} and \textit{false}
branches. Basically, this clause explicitly represents the semantics
of \textit{phi} and encodes both the data- and the control-flow
aspects of the execution, which allows OFC to handle faults in
both. Algorithm~\ref{alg:encodefunction} performs this encoding at
lines~\ref{alg:phi1}--\ref{alg:phi2}.

Finally, if \textit{st} is a traditional assignment statement, the
algorithm encodes \textit{st} as $st_{LHS} = st_{RHS}$, the
equivalence relation between the variable on \textit{st}'s lefthand
side and the expression on its righthand side (line
\ref{alg:normalassignment}). Because each assignment in SSA form
defines a new variable, $clause_{st}$ can be simply conjoined with the
current formula \textit{TF} (line~\ref{alg:conjoin}).

After processing a statement $st$ and generating the corresponding
clause $clause_{st}$, the algorithm records that $clause_{st}$ was
generated from $st$ and suitably updates the trace formula \textit{TF}
(lines~\ref{alg:updateco} and~\ref{alg:conjoin}). Finally, after
processing all statements in \textit{new\_trace},
\textit{FormulaGenerator} returns \textit{TF}.

  \begin{algorithm}[t]
 \begin{scriptsize}
 \DontPrintSemicolon
 \SetAlFnt{\scriptsize\scriptfont}
 \SetKwData{This}{this}
 \SetKwData{Up}{up}
 \SetKwFunction{Union}{Union}\SetKwFunction{FindCompress}{FindCompress}
 \SetKwInOut{Input}{Input}\SetKwInOut{Output}{Output}
 \caption{OFC}
 \label{alg:odin_main}
 \Input{
   \textit{ICFG}: ICFG of the faulty program\\
   \textit{TestSuite}: test suite for the program\\
 }
 \Output{
   faulty statements and their corresponding \textit{clauses}\\
 }
 \Begin{ 
    \textit{FIN} $\leftarrow$ GetFailingInput(\textit{TestSuite}) \; \label{alg:tc1} 
    \textit{ASSERT} $\leftarrow$ GetFailingAssertion(\textit{TestSuite}) \; \label{alg:tc2}
    \textit{TF} $\leftarrow  \{\}$ \;\label{alg:init1}
    \textit{SP} $\leftarrow \{\}$ \;
    \textit{clause\_origin} $\leftarrow \{\}$ \;
    \textit{visited\_branches} $\leftarrow  \{\}$ \;
    \textit{flip\_br} $\leftarrow$ null \;\label{alg:init2}
    \tcp{Step~1}
      \textit{new\_trace} $\leftarrow$ TraceGenerator(\textit{FIN}, \textit{visited\_branches}, \textit{flip\_br}) \label{alg:calltracegenerator}\;
      \textit{flip\_br} $\leftarrow$ null\label{alg:clearflipbr}\;
      \textit{TF} $\leftarrow$ FormulaGenerator(\textit{new\_trace}, \textit{TF}, \textit{ICFG}, \textit{SP}, \textit{clause\_origin})\label{alg:encode}\;
      \tcp{Step~2}
      \textit{CoMSSs} $\leftarrow$ Solver(\textit{FIN}, \textit{ASSERT}, \textit{TF})\label{alg:callsolver}\;
      \tcp{Step~3}
      \ForEach{\textit{CoMSS} in \textit{CoMSSs}\label{alg:resultanaylzer1}}{
        \ForEach{\textit{clause} in \textit{CoMSS}}{
          \textit{st} $\leftarrow$ \textit{clause\_origin(clause)}\;
          \If {\textit{st} is a conditional statement} {
              \label{alg:checkbranches}<\textit{true\_br}, \textit{false\_br}> $\leftarrow$ getBranches($st$)\;
              \If {\textit{visited\_branches(true\_br)}$==$null\label{alg:checktrue}} {
                \textit{flip\_br} $\leftarrow$ \textit{false\_br}\;
              go back to Step~1\label{alg:break1}\;
            }
            \If {\textit{visited\_branches(false\_br)}$==$null\label{alg:checkfalse}} {
                \textit{flip\_br} $\leftarrow$ \textit{true\_br}\;
              go back to Step~1\label{alg:break2}\;
            }
          }
        }
      \label{alg:resultanaylzer2}}

 \ForEach{\label{alg:wrapup1}\textit{CoMSS} in \textit{CoMSSs}} {
   \ForEach{\textit{clause} in \textit{CoMSS}} {
     report \textit{clause} and \textit{clause\_origin(clause)}\; \label{alg:finalreport}}
  \label{alg:wrapup2}}
}
\end{scriptsize}
\end{algorithm}

\subsubsection{Solver}
\label{sec:solver}

In its second step, OFC leverages a pMAX-SAT solver to find all
possible causes of the failure being considered. To do so, it invokes
function \textit{Solver} and passes to it the failing input, the
failing assertion, and the trace formula constructed in Step~1
(line~\ref{alg:callsolver} of Algorithm \ref{alg:odin_main}). Function
\textit{Solver} will first generate a formula by conjoining the input
clauses (\ie clauses that assert that the input is the failing input
\textit{FIN}), the current trace formula \textit{TF}, and the failing
assertion \textit{ASSERT}. Because \textit{FIN} causes the program to
fail, that is, to violate \textit{ASSERT}, the resulting formula is
unsatisfiable.

To suitably define the pMAX-SAT problem, \textit{Solver} encodes (1)
the input clauses and the failing assertion as hard clauses, (2) the
clauses in \textit{TF} generated from $\phi$ functions as hard
clauses, and (3) the other clauses in \textit{TF} as soft clauses.
The input clauses and the assertion are encoded as hard clauses
because the failure could be trivially eliminated by changing the
input or the assertion, which would not provide any information on
where the problem is in the program.  Encoding clauses generated by
$\phi$ functions as hard clauses, conversely, ensures that
control-flow related information is kept in the results, which is
necessary to handle control-flow related faults.  At this point,
function \textit{Solver} passes the so defined pMAX-SAT problem to an
external solver and retrieves from it all possible CoMSSs for the
problem (see Section~\ref{sec:bg}).

If CW were also used, OFC would generate a wpMAX-SAT problem instead
by assigning a weight to each soft clause based on the suspiciousness
of the corresponding program entity (\ie
\textit{clause\_origin(clause)}), as described in
Section~\ref{sec:weight-computation}.



\subsubsection{Result Analyzer}
\label{sec:result-analyzer}

OFC's third step takes the set of CoMSSs for the failure being
investigated, produced by Step~2, and generates a report with a set of
program entities (or an ordered list of entities, if we use CW and a
wpMAX-SAT solver) and corresponding clauses. The entities are
statements that, if suitably modified, would make the failing
execution pass (\ie the potential causes of the failure being
investigated). The expressions in the clauses associated with the
statements provide developers with additional information on how the
statements contribute to the failure, and as stated above, can thus be
seen as a mathematical explanation of the failure.

This part of OFC, corresponding to
lines~\ref{alg:resultanaylzer1}--\ref{alg:resultanaylzer2} of
Algorithm~\ref{alg:odin_main}, iterates through each clause of each
CoMSS computed in Step~2. For each clause, it first retrieves the
corresponding statement \textit{st}. If \textit{st} is a conditional
statement, the predicate in the conditional statement is potentially
faulty, and taking a different branch may fix the program. To account
for this possibility, the algorithm checks whether the conditional has
one branch that has not been executed in any previously computed trace
and, if so, expands the scope of the analysis by selecting that branch
as a new branch to analyze and going back to Step~1
(lines~\ref{alg:checkbranches}--\ref{alg:break2}). Step~1 would then
add such branch to the list of relevant branches, generate a new trace,
constructs a new formula, and perform an additional iteration of the
analysis.  Conversely, if both branches have already been covered, or
\textit{st} is not a conditional statement, the algorithm continues
and processes the next clause.


 \begin{algorithm}[t]
 \begin{scriptsize}
 \DontPrintSemicolon
 \SetAlFnt{\scriptsize\scriptfont}
 \SetKwData{This}{this}
 \SetKwData{Up}{up}
 \SetKwFunction{Union}{Union}\SetKwFunction{FindCompress}{FindCompress}
 \SetKwInOut{Input}{Input}\SetKwInOut{Output}{Output}
 \caption{TraceGenerator}
 \label{alg:tracegenerator}
 \Input{\textit{FIN}: failing input \\
   \textit{visited\_branches}: map from branches to traces that
   covered them \\
   \textit{flip\_br}: branch for which a new trace must be generated \\
 } \Output{
   \textit{new\_trace}: newly generated trace \\
 }
 \Begin{ 
     \eIf{\textit{flip\_br}$==$null}{
         \textit{new\_trace} $\leftarrow$ Execute(\textit{Input}, null, null)\label{alg:execute1}\;
     }{
         \textit{old\_trace} $\leftarrow$ \textit{visited\_branches(flip\_br)}\label{alg:execute2}\;
         \textit{new\_trace} $\leftarrow$ Execute(\textit{Input}, \textit{old\_trace}, \textit{flip\_br})\label{alg:execute3}\;
     }
     \ForEach{\textit{br} in \textit{new\_trace}}{\label{alg:insertbranch1}
     \If{\textit{visited\_branches(br)}==\textit{null}} {
       \textit{visited\_branches(br)} $\leftarrow$ \textit{new\_trace}}
     }\label{alg:insertbranch2}
     \Return{\textit{new\_trace}}\;
 }  
\end{scriptsize}
\end{algorithm}

If no clause in any CoMSS contains a conditional statement for which
one of the branches has not been covered, it means that the analysis
already considered the portion of the program relevant to the failure,
so the algorithm can terminate and produce a report
(lines~\ref{alg:wrapup1}--\ref{alg:wrapup2}). To do so, OFC iterates
once more through the set of CoMSSs computed during its last
iteration. 
For each clause in each CoMSS, OFC reports it to developers, together
with its corresponding statement, as a possible cause (and partial
explanation) of the failure.


 \begin{algorithm}[t]
 \begin{scriptsize}
 \DontPrintSemicolon
 \SetAlFnt{\scriptsize\scriptfont}
 \SetKwData{This}{this}
 \SetKwData{Up}{up}
 \SetKwFunction{Union}{Union}\SetKwFunction{FindCompress}{FindCompress}
 \SetKwInOut{Input}{Input}\SetKwInOut{Output}{Output}
 \caption{FormulaGenerator}
 \label{alg:encodefunction}
 \Input{
   \textit{ICFG}: ICFG of the faulty program \\
   \textit{TF}: current trace formula \\
   \textit{SP}: portion of the program currently considered \\
   \textit{new\_trace}: newly generated trace \\
 }
 \Output{\textit{TF}: updated trace formula\\
   \textit{SP}: updated portion of the program currently considered}
   \textit{clause\_origin}: map from clauses to statements that originated them \\
 \Begin{ 
     \ForEach{ \textit{st} $\in$ \textit{new\_trace}} {\label{alg:pass1_b}
         \If{\textit{st} $\notin$ \textit{SP}} {
             \textit{SP} $\leftarrow$ \textit{SP} $+$ \textit{st}\;
             \eIf{\textit{st} is a conditional statement} {
                 $predicate_{st} \leftarrow$ GetPredicate(\textit{st}) \label{alg:getpred}\;
                 $clause_{st} \leftarrow (guard_{st}=predicate_{st})$\label{alg:predassignment}\;
         }{
             \eIf{\textit{st} is a $\phi$ function}{
             \label{alg:phi1}
             \textit{phi} $\leftarrow \phi$ function in \textit{st}\label{alg:phicond1}\;
             \textit{cs} $\leftarrow \phi$'s conditional statement\;
             \textit{$guard_{cs}$} $\leftarrow$ \textit{cs}'s condition\label{alg:phicond2}\;
             $clause_{st} \leftarrow ((\textit{$guard_{cs}$} \wedge
             (st_{LHS}=st_{RHS,t})) \vee (\neg \textit{$guard_{cs}$} \wedge (st_{LHS}=st_{RHS,f})))$\;
     \label{alg:phi2} }{
         $clause_{st} \leftarrow (st_{LHS}=st_{RHS})$\label{alg:nonphi} \label{alg:normalassignment}\;
        
    }}
         \textit{clause\_origin}($clause_{st}$) = \textit{st}\label{alg:updateco}\;
         \textit{TF} $\leftarrow$ \textit{TF} $\wedge clause_{st}$\label{alg:conjoin}\; 
     }
  
 \label{alg:pass2_e}}
 \Return{\textit{TF}}\;
 }  
 \end{scriptsize}
\end{algorithm}
\DecMargin{1em}


\subsubsection{Illustrative Example}
\label{sec:illustrating-example}

\begin{figure}[t]
\centering
\includegraphics[width=.8\columnwidth,keepaspectratio]{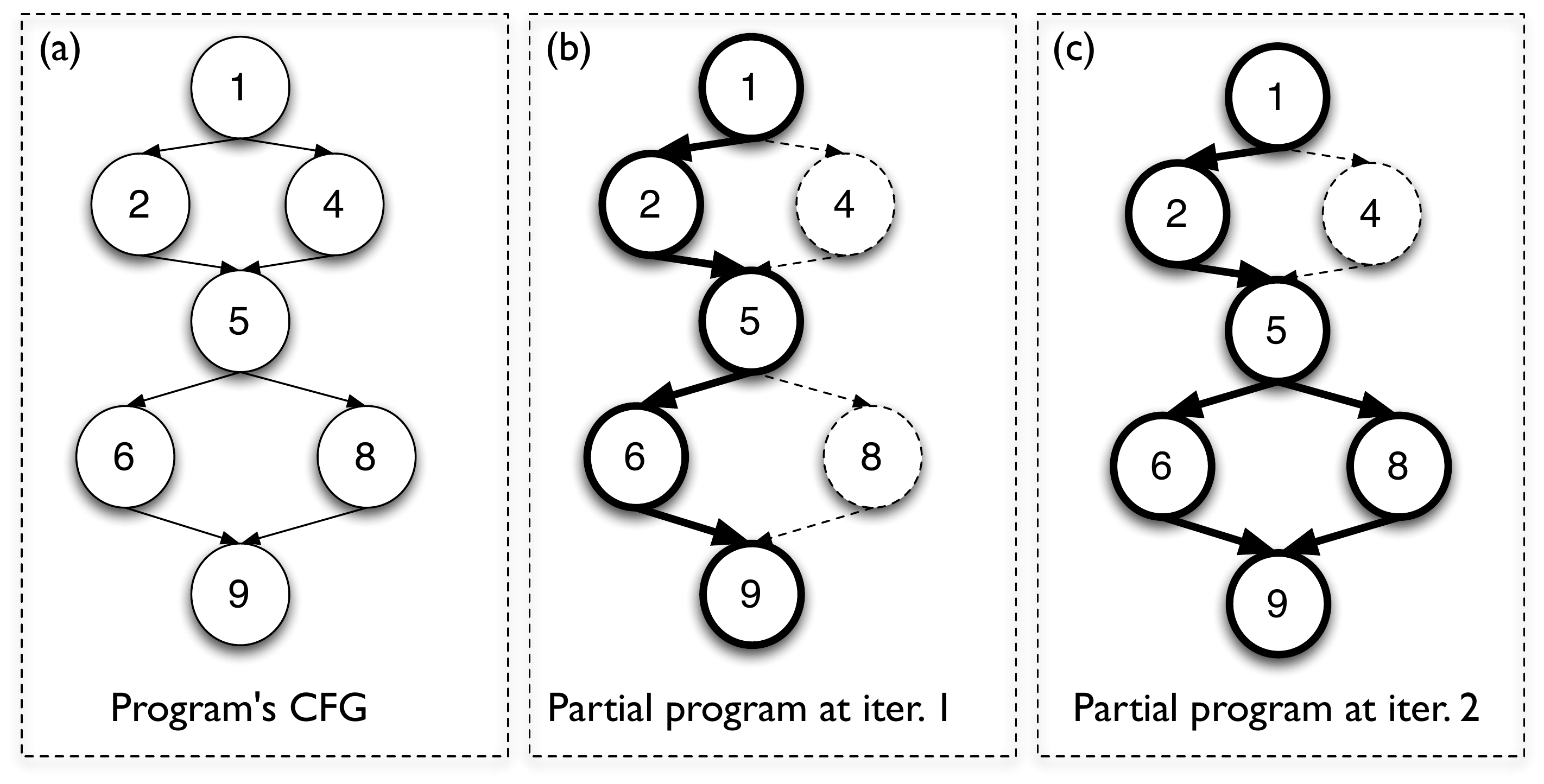}
\vspace{-12pt}
\caption{Control flow graph of \textit{P} (a) and partial \textit{P}
  considered during the first (b) and second (c) iteration of OFC.}
\label{fig:examplecfg}
\end{figure}

We recap how the different parts of OFC work together using a simple
program \textit{P} as an illustrative example.
Figure~\ref{fig:example1} shows P (left) and its SSA form (right),
whereas Figure~\ref{fig:examplecfg}(a) shows the \textit{P}'s Control
Flow Graph (CFG). \textit{P} takes two integer inputs and contains an
assertion at line~$9$. If we provide P with input {\scriptsize
  \{$x=0$, $y=0$\}}, the assertion is violated, as the execution
results in {\scriptsize $b=1$} and {\scriptsize $a=0$} at line~$9$.

This is the starting point of OFC: A faulty program in SSA form
(\textit{P}), a failing test case ({\scriptsize \{$x_{1}=0$,
  $y_{1}=0$\}}), and an assertion violated by the failing test case
({\scriptsize$b_{3} \le a_{3}$}).  Given these inputs, OFC operates in
three iterative steps.
%

In Step~1 of the first iteration, the \textit{Trace Generator} feeds
the failing input to \textit{P}, which results in the failing trace
\{$1$, $2$, $phi1$, $5$, $6$, $phi2$, $9$\}. This trace identifies the
partial program shown in Figure~\ref{fig:examplecfg}(b), where the
entities drawn in boldface are in the partial program, and those drawn
with dashed lines are ignored. (For simplicity, in the CFG we do not
show nodes corresponding to $\phi$ functions.) Given the generated
trace, the \textit{Formula Generator} computes a trace formula that
encodes the semantics of the partial program with respect to the
trace:

\vspace{-8pt}
\begingroup
\everymath{\scriptstyle}
\scriptsize
\begin{align*}
  TF_{1} = & (guard_{1}=(x_{1} \ge 0)) \wedge (a_{1}=x_{1}) \wedge \\
  &((guard_{1} \wedge (a_{3}=a_{1})) \vee (\neg guard_{1} \wedge (a_{3}=a_{2}))) \wedge\\
  & (guard_{2}=(y_{1} < 5)) \wedge (b_{1}=a_{3}+1) \wedge \\
  &((guard_{2} \wedge (b_{3}=b_{1})) \vee (\neg guard_{2} \wedge (b_{3}=b_{2})))
\end{align*}
\endgroup

In the trace, the second and fifth clauses correspond to statements
that do not involve $\phi$ functions. The first and fourth clauses
correspond to the predicates at lines $1$ and $5$ and contain two
extra variables, $guard_{1}$ and $guard_{2}$, that represent such
predicates. The third and sixth clauses represent the two statements
at lines $phi1$ and $phi2$, in which $\phi_1$ and $\phi_2$ define
$a_{3}$ and $b_{3}$. These clauses encode the information on which
variable a $\phi$ function may select and under which condition (\ie
the outcome of the guard), as described in
Section~\ref{sec:trace-gener-form}.

\begin{figure}[t]
\scriptsize
\centering
\begin{minipage}[c]{0.5\columnwidth}
\begin{alltt}
int {\bf P}(int x, int y) \{
1.  if (x>=0)
2.    a = x;
3.  else
4.    a = -x;

5.  if (y<5) 
6.    b = a+1;
7.  else 
8.    b = a+2;

9. assert(b<=a);
\}
\end{alltt}
\end{minipage}
\begin{minipage}[c]{0.4\columnwidth}
\begin{alltt}
   int {\bf P}(int x\(\sb{1}\), int y\(\sb{1}\)) \{
   1.  if (x\(\sb{1}\)>=0)
   2.    a\(\sb{1}\) = x\(\sb{1}\);
   3.  else
   4.    a\(\sb{2}\) = -x\(\sb{1}\);
phi1.  a\(\sb{3}\) = \(\phi\sb{1}\)(a\(\sb{1}\),a\(\sb{2}\));
   5.  if (y\(\sb{1}\)<5)
   6.    b\(\sb{1}\)=a\(\sb{3}\)+1;
   7.  else 
   8.    b\(\sb{2}\)=a\(\sb{3}\)+2;
phi2.  b\(\sb{3}\) = \(\phi\sb{2}\)(b\(\sb{1}\),b\(\sb{2}\));    
   9.  assert(b\(\sb{3}\)<=a\(\sb{3}\));
   \}
\end{alltt}
\end{minipage}
\vspace{-6pt}
\caption{Example code in normal (left) and SSA (right) form.}
\label{fig:example1}
\end{figure}

Step~2 of the algorithm then conjoins input conditions and failing
assertion with $TF_{1}$ to obtain the following complete,
unsatisfiable formula: {\scriptsize $CF_{1} = (x_{1}=0) \wedge
  (y_{1}=0) \wedge TF_{1} \wedge (b_{3} \le a_{3})$}


The algorithm now marks input clauses, failing assertion, and the two
clauses generated from $\phi$ functions as hard clauses, marks all
other clauses in $TF_1$ as soft clauses,
and feeds the result to a pMAX-SAT solver.  In this case, the solver
would return two CoMSSs: {\scriptsize $\{b_{1}=a_{3}+1\}$} and
{\scriptsize $\{guard_{2}=(y_{1}<5)\}$}.

When analyzing the set of clauses in all the CoMSSs, the third step of
the algorithm finds that there is one clause associated with a
conditional statement $c$ (the one at line~$5$, in this case). It thus
identifies the branches corresponding to $c$ (\ie branches $(5,6)$ and
$(5,8)$ in the CFG), and checks whether one of the branches was not
visited before. This is the case for branch $(5,8)$, so the algorithm
selects the unvisited branch as the branch to be expanded and returns
to Step~1.

In the second iteration of the algorithm, the \textit{Trace Generator}
re-executes \textit{P} with the same failing input, but forces the
execution to follow branch $(5,8)$ at conditional statement
$5$~\cite{tsankov2011execution}, which results in a new trace: \{$1$,
$2$, $phi1$, $5$, $8$, $phi2$, $9$\}.  Given this trace, the algorithm
first adds the newly covered program entities to the partial program
(see Figure~\ref{fig:examplecfg}(c)) and then computes a new trace
formula based on the expanded partial program. Since the execution of
statement $8$ instead of statement $6$ is the only difference between
this trace and the one generated in the previous iteration, the new
trace formula is identical to $TF_1$, except for clause {\scriptsize
  $(b_{1}=a_{3}+1)$} (corresponding to statement $6$), which is
replaced by clause {\scriptsize $(b_{2}=a_{3}+2)$} (corresponding to
statement $8$).  This trace formula, when conjoint with $TF_1$, would
thus result in the trace shown below. (In practice, OFC simply
conjoins the previous formula and the clause(s) corresponding to the
new statement(s), which produces the same result, as explained in
Section~\ref{sec:trace-gener-form}.)

\vspace{-8pt}
\begingroup
\everymath{\scriptstyle}
\scriptsize
\begin{align*}
  TF_{2} = &  (guard_{1}=(x_{1} \ge 0)) \wedge (a_{1}=x_{1}) \wedge \\
  &((guard_{1} \wedge (a_{3}=a_{1})) \vee (\neg guard_{1} \wedge (a_{3}=a_{2}))) \wedge\\
  & (guard_{2}=(y_{1} < 5)) \wedge  (b_{1}=a_{3}+1) \wedge \\
  &((guard_{2} \wedge (b_{3}=b_{1})) \vee (\neg guard_{2} \wedge
  (b_{3}=b_{2}))) \wedge (b_{2}=a_{3}+2)
\end{align*}
\endgroup

Similar to the first iteration, the algorithm then conjoins input
conditions ($TF_{2}$) and failing assertion to obtain a new
unsatisfiable formula, marks hard and soft clauses, and
feeds the formula to the pMAX-SAT solver, which would return two
CoMSSs: {\scriptsize $\{b_{1}=a_{3}+1\}$} and {\scriptsize
  $\{guard_{2}=(y_{1} < 5), b_{2}=a_{3}+2\}$}.

Step~3 of the algorithm then checks whether any of these clauses is
associated with a conditional statement $cs$ and, if so, whether $cs$
has any outgoing branches not yet visited by a trace. In this case,
both of the branches corresponding to the first clause in the second
CoMSS have been covered in our analysis. Therefore, the algorithm
stops iterating and reports to developers these two CoMSSs, together
with their corresponding program entities: $\{$line~$6\}$ and
$\{$line~$5$, line~$8\}$.

This would inform developers that suitably changing either (1) the
statement at line~$6$ or (2) both the conditional statement at
line~$5$ and the statement at line~$8$ could fix the program, so that
input ({\scriptsize $x_{1}=0$, $y_{1}=0$}) would not violate the
assertion at line~$9$. The clauses associated with the statements
would provide additional information that could help understand the
fault and find a fix---a (partial) mathematical explanation of how the
statements contribute to the failure.

\subsubsection{Further Considerations}
\label{sec:considerations}

Compared to BugAssist, OFC tends to generate a simpler formula that is
as effective as one that encodes the whole program but less expensive
to solve. Considering our example, for instance, OFC explored only $2$
of the $4$ paths in the program. Compared to an all-paths analysis,
OFC included only \textit{relevant} program entities in the trace
formula: the assertion violation in the example is independent from
the outcome of the predicate at statement~$1$, and our algorithm
successfully identifies the statement as irrelevant and avoids
exploring both of its branches.
However, although we expect the cost of finding solutions for a
formula constructed by OFC to be lower than that of solving the
formula generated by an all-paths analysis, OFC can perform a number
of iterations when constructing the formula, and thus make multiple
calls to the solver. Therefore, whether OFC is more efficient than an
all-paths analysis depends on the number and cost of iterations it
performs. We study this tradeoff in our empirical evaluation of OFC
(see Section~\ref{sec:rq2-ondemand}).

Compared to approaches that consider only the failing trace (\eg
\cite{Ermis2012, Christ13}), OFC can conservatively identify all parts
of the program that are relevant to the failure. In our example, if
the algorithm had stopped after the first iteration, it would have
missed the second CoMSS: {\scriptsize$\{guard_{2}=(y_{1}<5),
  b_{2}=a_{3}+2\}$}. That is, by reasoning about the original failing
trace alone, developers could only infer that the fault may be related
to the conditional statement at line~$5$. Conversely, by considering
also the additional trace, OFC can discover that a fix involving that
conditional statement should also consider possible changes to
statement $8$. OFC can thus make formula-based debugging more
efficient without losing accuracy and effectiveness.

Compared to more traditional debugging techniques, OFC is likely to
produce more accurate results. If we applied dynamic slicing to our
example, for instance. A dynamic slice computed for the failing
assertion at line~9 would include not only statements $5$ and $6$,
which is correct, but also statements $1$ and $2$, which are
irrelevant for the failure.
%
%

\subsubsection{Additional Details and Optimizations}
\label{sec:furth-optim}

\paragraph*{Handling Multiple Failing Inputs}

Although it is defined for a single failing input, OFC can take
advantage of the presence of multiple failing tests \textit{for the
  same fault}. Because, by definition, the faulty statement(s) should
be executed by all failing tests and be responsible for all observed
failures, OFC can handle multiple failing inputs as follows: (1)
generate a report for each individual failing input, (2) identify the
potential faulty entities (and corresponding clauses) that appear in
all individual reports, and (3) report to developers only these
entities, ranked based on the average of their original ranks in the
individual reports.



\paragraph*{Loop Unrolling}

In the presence of loops, the size of a formula is in general
unbounded.  As it typical for symbolic analysis approaches (\eg
\cite{ckl2004, Merz2012, Jose2011}), in OFC we address this issue by
performing loop unrolling~\cite{dragonbook2006}. One advantage of OFC
over other all-paths analyses is that it can decide how many times to
unroll a given loop based on concrete executions, rather than on some
arbitrary threshold. Nevertheless, for practicality reason, OFC still
needs to define an upper bound for loop unrolling, to limit the
overall size of trace formulas.

\paragraph*{Dynamic Symbolic Execution}

OFC, like dynamic symbolic execution~\cite{sen05, godefroid05}, may
replace symbolic variables in the trace formula with their
corresponding concrete values, so as to allow the solver to handle
formulas that go beyond its theories (\eg non-linear expressions,
dynamic memory accesses). Doing so makes the approach more practical,
but can introduce unsoundness (in the form of discrepancies between
the actual semantics of the program and the semantics encoded in the
formula) and reduce the number of possible solutions the solver can
compute. This can result in both false positives---program entities
that, even if suitably changed, could not eliminate the failure at
hand---and false negatives---solutions that do not include the faulty
statement(s).

\paragraph*{Solution Space Pruning}

Because the number of CoMSSs for a given MAX-SAT problem may be too
large and affect the ability of OFC to enumerate and analyze all
solutions in a reasonable amount of time, OFC allows developers to
specify an upper bound for the number of clauses in a CoMSS (\ie the
number of statements reported \textit{together as a single fault}) and
terminates the search for new solutions when the solver starts
reporting CoMSSs that exceed this bound.  The rationale is that a
potential bug generally involves a limited number of statements,
whereas a CoMSS that contains a large number of clauses suggests a
large semantic change in the program (which may be able to eliminate a
failure but is usually not an ideal fix).

\section{Empirical Evaluation}
\label{sec:empir-invest}

To evaluate CW and OFC, we have developed a prototype tool for C
programs that implements four different formula-based debugging
techniques: BugAssist (BA), BugAssist with clause weighting (BA+CW),
on-demand formula computation (OFC), and on-demand formula computation
with clause weighting (OFC+CW). We have then empirically investigated
the following research questions:

\begin{itemize}\denseitems

\item \textbf{RQ1:} Does BA+CW produce more accurate results than BA?
  If so, what is CW's effect on efficiency?

\item \textbf{RQ2:} Does OFC improve the efficiency of an all-paths
  formula-based debugging technique?

\item \textbf{RQ3:} Does OFC+CW combine the benefits of OFC and CW? If
  so, can it scale to programs that an all-paths technique could not
  handle?
\item \textbf{RQ4:} How dependent are our results on the specific
  solver used?



\end{itemize}

\noindent
We now discuss our evaluation setup and our results.

\subsection{Evaluation Setup}
\label{sec:evaluation-setup}

\paragraph*{Implementation}

We implemented OFC, as presented in Section~\ref{sec:algo}, in Java
and C. Our tool leverages the LLVM compiler infrastructure ({\small
  \url{http://llvm.org/}}) to transform programs into SSA form and add
instrumentation that (1) dumps dynamic traces and concrete program
states and (2) performs execution
hijacking~\cite{tsankov2011execution} to force the program along
specific branches in the \textit{Trace Generator}.  We implemented BA
as a version of OFC that builds a formula for all (bounded) paths in
the program instead of operating on demand.
We implemented Ochiai~\cite{ochiai}, the statistical fault
localization technique that we use for CW, as a Java program that
operates directly on the dumped dynamic traces.
%
Finally, to handle wpMAX-SAT and pMAX-SAT problems, we implemented
interfaces to invoke the Yices SMT solver~\cite{yices} and
the Z3 theorem prover~\cite{z3}. 
We used the Yices solver for the first three research questions, as Z3
does not provide wpMAX-SAT capabilities.

Implementing the OFC algorithm, and in particular the \textit{Trace
  Generator} and the \textit{Formula Generator} components, is
extremely challenging both from the conceptual and the engineering
standpoint~\cite{ckl2004}. To avoid spending too much development
effort, we decided to build a prototype that implements OFC
completely, but has some limitations when handling some constructs of
the C language related to heap memory management.

\paragraph*{Benchmarks}

For our evaluation, we selected three benchmarks. The first two
benchmarks consist of multiple (faulty) versions of two programs in
the SIR repository~\cite{sirrepo}: \textsf{tcas} ($41$ versions,
\textasciitilde{}200 LOC) and \textsf{tot\_info} ($11$ versions,
\textasciitilde{}500 LOC). These programs also come with test cases
and a golden (supposedly fault-free) version that can be used as an
oracle. We selected these programs for two main reasons.  First,
\textsf{tcas} is an ideal subject for our evaluation because it allows
us to find all possible solutions of the program formulas considered
and thus precisely compute the savings that OFC achieves in terms of
complexity reduction. (This is in general impossible for larger, more
complex programs.) Second, these two programs were also used to
evaluate BugAssist~\cite{Jose2011}, which lets us directly compare our
results with those of a state-of-the-art all-paths formula-based
technique in terms of accuracy and efficiency.  The third benchmark we
considered is a faulty version of \textsf{Redis}, a widely used
in-memory key-value database (\textasciitilde{}32 KLOC), which also
comes with a set of test cases.

\paragraph*{Study Protocol}

For each faulty program version considered, we proceeded as follows.
\textit{First}, we identified passing and failing test cases for that
version. For \textsf{tcas} and \textsf{tot\_info}, we did so by
defining the assertion for a test using the output generated by the
same test when run against the golden implementation. For the bug in
\textsf{Redis}, we used the bug description~\cite{redisbug} and the
corresponding test~\cite{redis-test}. We then ran all programs
instrumented to collect coverage information for all passing and
failing tests at the same time.  We used this coverage information to
compute the suspiciousness values for the branches and statements in
each program version using the Ochiai metric~\cite{ochiai}.  These are
the values that BA+CW and OFC+CW use to assign weights to the clauses
in the program formula. \textit{Second}, for each failing input, we
ran all four techniques on the faulty version. Because the all-paths
techniques timed out or could not build a formula for the bugs in
\textsf{tot\_info} and \textsf{Redis} (see
Section~\ref{sec:rq2-effect-our} for details), we could only
investigate RQ1 and RQ2 on \textsf{tcas}, whereas we used all three
benchmarks for RQ3. (For fairness, we note that
Reference~\cite{Jose2011} reports results for 2 versions of \textsf{tot\_info}.
However, the authors mention that those results were obtained working
on a program slice, and there are no details on how the slice was
computed and on which version, so we could not replicate them using
either our or their implementation of BA.)
For faults with multiple failing test cases, in each technique, we
combined the results for the individual inputs, so as to generate a
final report for each faulty version and for each technique.
For each faulty version and for each technique that successfully ran
on it, the technique generated a report for the developers. To do a
complete assessment of the performance of the techniques, we also
recorded the average CPU time of each technique for each failing
input, the number of iterations of the OFC algorithm, whether the
generated report contained the fault, and, if so, the rank of the
fault in the report.

\setlength{\tabcolsep}{3pt}
\subsection{Results and Discussion}

\subsubsection{RQ1---BA+CW Versus BA}
\label{sec:rq1-weighted}

To answer this research question, we compared the accuracy and the
computational cost of BA+CW and BA to evaluate the impact of
leveraging information from statistical fault localization. To do so,
we ran both techniques on the $41$ faulty versions of \textsf{tcas}
and computed the results as described in
Section~\ref{sec:evaluation-setup}. Table~\ref{tab:fl_comp_1} presents
these results.  The columns in the table show the version ID, the
number of lines of code a developer would have to examine before
getting to the fault, and the average CPU time consumed by BA and
BA+CW to compute their results. For comparison purposes, in the last
column we also report the results of a traditional fault-localization
technique (Ochiai). For \textsf{tcas.v3}, for instance, it took $292$
seconds (BA) and $183$ seconds (BA+CW) to generate the results, and
developers would have to examine $8.5$ lines of code (BA), $1$ line of
code (BA+CW), or 3 lines of code (Ochiai). Note that, for BA+CW, the
number of lines of code to examine corresponds to the actual rank of
the faulty line of code in the report produced by the technique. BA,
however, does not rank the potentially faulty lines of code, but
simply reports them as an unordered set to developers.  Therefore, the
number in the table corresponds to the number of lines of code
developers would have to investigate if we assume they examine the
entities in the set in a random order (\ie half of the size of the
set).

\begin{table}[t]
  \caption{Results for BA and BA+CW when run on \textsf{tcas}.}
\vspace{-7pt}
\label{tab:fl_comp_1}
\centering
\begin{scriptsize}
\begin{tabular}{|l|l|l|l|l|l||l|l|l|l|l|l|}
  \hline
  \tiny Version & \multicolumn{2}{c|}{\scriptsize BA} & \multicolumn{2}{c|}{\scriptsize BA+CW}& \tiny Ochiai&
  \tiny Version & \multicolumn{2}{c|}{\scriptsize BA} & \multicolumn{2}{c|}{\scriptsize BA+CW}& \tiny Ochiai\\
  \hline
  & rank & time & rank & time & rank && rank & time &rank & time& rank\\
  \hline
  v1  & 7.5 & 26s & 2& 27s & 4 &v22 & 4  & 7s & 5& 7s&22 \\
  \hline
  v2 & 4 & 15s & 4 & 16s & 3 & v23 &  5.5 & 15s & 10 &12s&23 \\
  \hline
  v3  & 8.5 & 292s & 1 & 183s & 3&v24 &  7.5 & 30s & 8 & 23s&23\\
  \hline
  v4  &  8& 11s & 3 & 11s & 1& v25 &  5.5 & 297s & 4 &216s &2\\
  \hline
  v5  & 7.5 & 352s &  3& 323s & 18& v26 &  8 & 160s & 5 & 123s&21 \\
  \hline
  v6 & 7.5 & 569s& 5 & 316s & 4& v27 &  9.5 & 443s& 4& 393s & 21\\
  \hline
  v7 & 8 & 484s & 8 & 238s & 8& v28 &  5 & 41s & 3 & 40s& 2\\
  \hline
  v8 & 7.5 & 21s & 13& 18s & 48& v29 &  5 & 25s & 1 & 27s&20\\
  \hline
  v9  & 4.5 & 18s & 10 & 15s & 23& v30 &  5 & 11s & 6 & 14s & 20\\
  \hline
  v10  & 8 & 125s & 3 &96s & 4 &v31 &  8.5  & 958s & 2& 909s&4\\
  \hline
  v11  & 5.5 & 130s & 1 & 91s & 21& v32 &  8.5 & 171s & 1 &145s& 3\\
  \hline
  v12 & 8 & 22s & 11 & 20s & 49 & v33 &  6 & 79s &1  & 70s& 3\\
  \hline
  v13  & 8 & 24s & 7 & 21s & 1 &v34 &  7.5 & 164s & 5 & 144s&23 \\
  \hline
  v14  & 7 & 28s & 1 & 28s &1& v35 &  5 & 38s & 3 & 40s& 2\\
  \hline
  v15  & 6.5 & 14s & 5 & 14s & 21& v36 &  2.5 & 19s &1 & 17s&1 \\
  \hline
  v16 &  8 & 331s & 12 & 228s &49 & v37 &  7.5 & 127s & 1 & 136s& 3 \\
  \hline
  v17  & 8 & 626s & 8 & 285s &49& v38 &  6.5 & 8s & 1& 8s & 2\\
  \hline
  v18  & 6 & 378s & 6 & 245s & 49& v39 & 6 & 244s & 4 & 272s & 2\\
  \hline
  v19  & 8 & 399s & 5 & 167s & 49&v40 & 5.5 & 219s & 3 & 219s & 4\\
  \hline
  v20  & 8 & 504s & 8 & 247s & 21&v41 &  7.5 & 6s & 2 & 5s& 6\\
  \hline
  v21  & 7.5 & 252s & 8 & 194s & 21 & \tiny Average & 6.5 & 187s & 4.7 & 137s& 17\\
  \hline
\end{tabular}
\end{scriptsize}

\end{table}


As the results in Table~\ref{tab:fl_comp_1} show, both techniques were
able to identify the faulty statements for all versions considered.
We can also observe that both BA and BA+CW produced overall more
accurate results that Ochiai (significance level of 0.05 for both BA
and BA+CW for \emph{a paired t-test}). Although this was not a goal of the
study, it provides evidence that formula-based techniques, by
reasoning on the semantics of a failing execution, can provide more
accurate results than a purely statistical approach. As for the
comparison of BA and BA+CW, BA+CW produced better results than BA,
with a significance level of 0.05 for \emph{a paired t-test}. On average, a
developer would have to examine $4.7$ statements per fault for BA+CW
versus $6.5$ for BA. By leveraging the suspiciousness values computed
by statistical fault localization, BA+CW can thus outperform BA in
most cases ($33$ out of $41$). For the $8$ cases in which BA+CW did
not outperform BA, manual analysis of the results identified one main
reason.  In some cases, the weights computed by fault localization
were too inaccurate and caused the solver to first produce CoMSSs that
did not include the actual faulty statements. Despite these negative
cases, the overall performance of BA+CW is remarkable and justify the
use of statistical fault-localization information. BA+CW ranked the
faulty statement first for $9$ out of $41$ versions, among the top $3$
statements in another $8$ cases, and at a position greater than $10$
in only $3$ cases.

The data in Table~\ref{tab:fl_comp_1} also allow us to investigate the
second part of RQ1, that is, the effect of CW on efficiency. As we
discussed in Section~\ref{sec:weight-computation}, solving wpMAX-SAT
problems may be computationally more expensive than solving a pMAX-SAT
problem, so the use of CW may negatively affect the efficiency of
formula-based debugging. As the table shows, on average BA+CW performs
significantly better than BA (137s versus 187s, significance level of
0.05). Although these results may seem counterintuitive, we discovered
that the extra information provided by the weights can in many cases
unintentionally help the solver find CoMSSs more efficiently.

In summary, our results provide initial evidence that CW can improve
formula-based debugging, both in terms of accuracy and in terms of
efficiency.

\setlength{\tabcolsep}{2pt}
\begin{table}[t]
  \caption{Performance results for BA and OFC on \textsf{tcas}.}
\vspace{-7pt}
\label{tab:ondemand}
\centering
\begin{scriptsize}
\begin{tabular}{|l|l|l|l|l||l|l|l|l|l|}
  \hline
  \tiny Version & \tiny BA & \tiny OFC &  \tiny \textit{\#Iteration}  & \tiny \textit{Time per iteration} & 
  \tiny Version & \tiny BA & \tiny OFC & \tiny \textit{\#Iteration} & \tiny \textit{Time per iteration}   \\
  \hline
  v1  & 26s & 7s & 9 & 0.8s & v22 & 7s & 6s & 13.2 & 0.4s \\
  \hline
  v2 & 15s & 38s & 12 & 3.2s & v23 & 15s & 24s & 11 & 2.1s \\
  \hline
  v3 & 292s & 19s & 14 & 1.4s &v24 &  30s & 7s & 10 & 0.7s\\
  \hline
  v4  & 11s & 6s & 9.2 & 0.6s & v25 &   297s & 244s & 12 & 20.3s \\
  \hline
  v5  &  352s & 15s & 13.4 & 1.1s & v26 &   160s & 17s & 13 & 1.3s \\
  \hline
  v6 & 569s& 17s & 13 & 1.3s & v27 &   443s& 15s & 13.4 & 1.1s\\
  \hline
  v7 & 484s & 104s & 14.8 & 7.1s & v28 &   41s & 24s & 11.2 & 2.2s\\
  \hline
  v8 & 21s & 5s & 10 & 0.5s & v29 &   25s & 6s & 9.8 & 0.6s\\
  \hline
  v9 & 18s & 28s & 12 & 2.4s &  v30 & 11s & 24s & 11 & 2.2s\\
  \hline
  v10  & 125s & 22s & 14 & 1.6s& v31  & 958s & 33s & 10.8 & 3s\\
  \hline
  v11 & 130s & 11s & 8.4 & 1.3s & v32  & 171s & 14s & 13 & 1.1s\\
  \hline
  v12 & 22s & 17s & 14.2 & 1.2s & v33 & 79s &178s & 13 & 13.7s\\
  \hline
  v13  & 24s & 15s & 13.3 & 1.2s &  v34 & 164s & 21s & 13 & 1.6s \\
  \hline
  v14  & 28s & 20s & 13.8 & 1.4s & v35  & 38s &22s & 14 & 1.5s\\
  \hline
  v15  & 14s & 20s & 13.2 & 1.5s & v36 & 19s &11s & 11.2 & 1s \\
  \hline
  v16  & 331s & 16s & 13 & 1.2s & v37  & 127s & 251s & 14 & 18s \\
  \hline
  v17  & 626s & 73s & 14.2 & 5.1s & v38  & 8s & 95s & 16 & 5.9s \\
  \hline
  v18  & 378s & 96s & 13.4 & 7.2s & v39  & 244s & 213s & 12 & 17.8s \\
  \hline
  v19  & 399s & 17s & 13.2 & 1.3s & v40  & 219s & 180s & 10.4 & 17.3s \\
  \hline
  v20  & 504s & 7s & 9.4 & 0.8s &v41  & 6s & 5s & 8.2 & 0.6s\\
  \hline
  v21  & 252s & 6s & 8.8 & 0.7s & \tiny Average & 187s & 48s & 12 & 4s\\
  \hline
\end{tabular}
\end{scriptsize}
\end{table}

\subsubsection{RQ2---OFC Versus BA}
\label{sec:rq2-ondemand}

To investigate RQ2, we compared OFC and BA in terms of efficiency.  As
we did for RQ1, we ran the two techniques on the $41$ faulty versions
of \textsf{tcas} and measured their performance. The results are shown
in Table~\ref{tab:ondemand}. The table shows the version ID, the
average CPU time spent by BA and OFC, respectively, on each failing
input, the number of iterations (\ie path expansions) of the OFC
algorithm, and the average CPU time spent by OFC in each iteration.
For example, for a failing input in \textsf{tcas.v1}, it took $26$
seconds (BA) and $7$ seconds (OFC) to generate the results, OFC
iterated 9 times, and, for each expansion, it took OFC $0.8$ seconds
to find all CoMSS solutions.

The second and fifth columns in the table clearly show that it took
considerably less time for the pMAX-SAT solver to find solutions for
formulas generated in one iteration of OFC (4 seconds) than for
formulas generated by BA (187 seconds). The statistically significant
gain of efficiency (significance level of 0.05) is caused, as
expected, by the difference in the complexity of the encoded
formulas---OFC only encodes the subset of the program relevant to the
failure into the formulas passed to the solver, while BA generates a
much more complex formula that encodes the semantics of the entire
program.

The results in the fourth column of Table~\ref{tab:ondemand} indicate
that OFC performed 12 iterations per fault, on average.  Therefore, as
we discussed in Section~\ref{sec:considerations}, the benefits of
generating a simpler formula were in some cases (\eg \textsf{tcas.v2})
outweighed by the cost of solving multiple pMAX-SAT problems during
on-demand expansion, thus making OFC less efficient than BA. In fact,
comparing the results in the second and third columns of the table, we
can observe that there were $8$ cases in which OFC performed worse
than BA.

Overall, however, OFC was more efficient than BA in $33$ out of $41$
cases and could achieve almost 4X speed-ups on average (48 versus 187
seconds) and over 70X speed-ups in the best case (504 versus 7
seconds).  Also in this case, the difference in performance between
the two techniques was statistically significant at the 0.05 level.

It is also worth noting that our results on the number of iterations
performed by OFC provide some evidence that techniques that operate on
a single-trace formula (\eg \cite{Ermis2012, Christ13}) may compute
inaccurate results, even when they encode both data- and control-flow
information. Because each expansion adds new constraints that must be
taken into account in the analysis, limiting the analysis to a single
trace is likely to negatively affect the quality of the results.

Finally, as a sanity check, we examined the sets of suspicious
entities reported by the two techniques. This examination confirmed
that OFC reports the same sets as BA (\ie the fault-ranking results
for OFC were the same as those for BA, shown in
Table~\ref{tab:fl_comp_1}).  That is, it confirmed that OFC is
able to build smaller yet conservative formulas and can thus produce
the same result as an approach that encodes the whole program.

In summary, our results for RQ2 provide initial, but clear evidence
that OFC can considerably improve the efficiency of formula-based
debugging without losing effectiveness with respect to an all-paths
technique such as BugAssist.

\begin{table}[t]
  \caption{Average time for processing \textsf{tcas} faults.}
\label{tab:combined}
\centering
\vspace{-7pt}
\begin{scriptsize}
\begin{tabular}{|l|l|l|l|}
  \hline
  \scriptsize BA & \scriptsize BA+CW & \scriptsize OFC & \scriptsize OFC+CW\\
  \hline
  187s & 137s & 48s & 36s \\
  \hline
\end{tabular}
\end{scriptsize}
\end{table}

\subsubsection{RQ3---OFC+CW Versus BA, BA+CW, and OFC}
\label{sec:rq2-effect-our}

To answer the first part of RQ3, we compared the performance of OFC+CW
with that of the other three techniques considered, in terms of both
accuracy and efficiency, when run on the $41$ \textsf{tcas} versions.
For accuracy, we found that the results for OFC+CW, not reported here
for brevity, were the same as those listed in the ``BA+CW'' column of
Table~\ref{tab:fl_comp_1}. This is not surprising, as OFC reports the
same sets as BA, as we just discussed, and we expect CW to benefit
both techniques in the same way. Therefore, the results show that
OFC+CW is as accurate as BA+CW and more accurate than BA and OFC.

To compare the efficiency of the four techniques considered, we
measured the average CPU time required by the techniques to process
one fault in \textsf{tcas}, shown in Table~\ref{tab:combined}. As the
table shows, for the cases considered, combining OFC and CW can
further reduce the cost of formula-based debugging by 25\% with
respect to OFC and by over 80\% with respect to our baseline, BA.
Although these are considerable improvements, it is unclear whether they can actually
result in more scalable formula-based debugging. This is the focus of
the second part of RQ3, which aims to assess the potential increase in
scalability that our two improvements can provide.

To answer this part of RQ3, we ran the techniques considered on our
other two benchmarks: \textsf{tot\_info} and \textsf{Redis}.

\paragraph*{\textsf{tot\_info} results for RQ3}

Unlike \textsf{tcas}, \textsf{tot\_info} contains loops, calls to
external libraries, and complex floating point computations. (We
considered all faults except those directly related to calls to
external \textit{system} libraries, which our current implementation
does not handle.) Because of the presence of loops, and as discussed
in Section~\ref{sec:furth-optim},
we set an upper bound of 5 to the size of clauses in a CoMSS.  (We
believe 5 is a reasonable value, as it means that the technique would
be able to handle all faults that involve up to 5 statements.)  As we
discussed in Section~\ref{sec:evaluation-setup}, for BA and BA+CW the
program formula generated was too large, and the solver was not able
to compute the set of CoMSSs within two hours (the time limit we used
for the study) for the faults considered.  Conversely, OFC and OFC+CW
were able to compute a result within the time limit for all faults,
which provides initial evidence that our improvements can indeed
result in more scalable formula-based techniques. By focusing only on
the relevant parts of a failing program and leveraging statistical
fault localization, OFC+CW can reduce the complexity of the analysis
and successfully diagnose faults that an all-paths technique may not
be able to handle.  To also assess the accuracy of the produced
results, in Table~\ref{tab:fl_totinfo} we show the results computed by
OFC+CW. The columns in the table show the program version and the
number of lines of code a developer would have to examine before
getting to the fault in that version.  As the table shows, OFC+CW was
able to rank all $11$ faults within the top $10$ statements in the
list reported to the developer, and $4$ of them at the top of the
list.

\begin{table}[t]
    \caption{Ranking results of OFC+CW on \textsf{tot\_info}}
\vspace{-7pt}
\label{tab:fl_totinfo}
\centering
\begin{scriptsize}
\begin{tabular}{|l|l||l|l||l|l|}
  \hline
  \scriptsize Version&  \scriptsize OFC+CW &  \scriptsize Version& \scriptsize OFC+CW &  \scriptsize Version& \scriptsize OFC+CW\\
  \hline
  tot\_info.v1 & 2 & tot\_info.v14 & 1 & tot\_info.v20 & 3 \\
  \hline
  tot\_info.v3 & 1 & tot\_info.v15 & 1 & tot\_info.v22 & 6 \\
  \hline
  tot\_info.v4 & 1 &  tot\_info.v16 & 2 &  tot\_info.v23 & 8 \\
  \hline
  tot\_info.v11 & 3 & tot\_info.v18 & 3 & & \\
  \hline
\end{tabular}
\end{scriptsize}

\end{table}


\begin{figure}[t]
\tiny
\centering
\begin{minipage}[c]{0.8\columnwidth}
\begin{alltt}
203 #define LUA_CMD_OBJCACHE_SIZE 32
...
206 int j, argc = lua_gettop(lua);
...
214 static robj *cached_objects[LUA_CMD_OBJCACHE_SIZE];
...
218 if (argc == 0) {
...
221  return 1;
222 }
...
232  for (j = 0; j < argc; j++) \{
233    char *obj_s;
234    size_t obj_len;
236    obj_s = (char*)lua_tolstring(lua,j+1,&obj_len);
237    if (obj_s == NULL) break; /* Not a string. */
    /* Try to use a cached object. */
    /* bug fixes */
240-   if (cached_objects[j] && cached_objects_len[j] >= obj_len) \{
240+   if (j < LUA_CMD_OBJCACHE_SIZE && cached_objects[j] && 
241+     cached_objects_len[j] >= obj_len) \{
...
\end{alltt}
\end{minipage}
\caption{Excerpt code of the bug in Redis.}
\label{fig:redisbug}
\end{figure}

\begin{table}[t]
  \caption{Results for OFC+CW when run on the bug from \textsf{Redis}.}
\vspace{-7pt}
\label{tab:redis_result}
\centering
\begin{scriptsize}
\begin{tabular}{|l|l|l|}
  \hline
  Rank & Source Location & Statement\\
  \hline
  1 & scripting.c:237 & if (obj\_s == NULL) break; \\
  \hline
  2 & scripting.c:236 & obj\_s = (char*)lua\_tolstring(lua,j+1,\&obj\_len);\\
  \hline
  3 & scripting.c:232 & j++ \\
  \hline
  4 & scripting.c:206 & int j, argc = lua\_gettop(lua);\\
  \hline
  5 & scripting.c:218 & if (argc == 0) \\
  \hline
\end{tabular}
\end{scriptsize}
\end{table}

\paragraph*{\textsf{Redis} results for RQ3}

To further assess the scalability of OFC+CW, we run the techniques
considered on our third benchmark, a real-world bug~\cite{redisbug} in
\textsf{Redis}, which is considerably larger and more complex than
\textsf{tcas} and \textsf{tot\_info}.  The bug is a potential buffer
overflow in a module of \textsf{Redis} that processes Lua scripts
({\small \url{www.lua.org/}}) and consists of ~1 KLOC).  Figure
\ref{fig:redisbug} shows an excerpt of the bug.  The original version
of the code fails to check whether the size of the script from the
command line is greater than the size of the memory in which it is
stored. If the script is too large, the program generates an
out-of-boundary memory access and fails.

We inserted assertions that are triggered when a buffer overflow
occurs, and applied OFC+CW to the faulty code. Our tool generated the
report shown in Table~\ref{tab:redis_result}, which contains five
suspicious statements and program locations. The first entry in our
report suggests that a control statement should be changed after line
237 of scripting.c to avoid the out-of-boundary access in the next
statement. This is also the location where the developers of
\textsf{Redis} fixed the bug~\cite{redisbug}. Also in this case, we
tried to run the all-paths techniques on the module, but they were not
successful. Because BA relies on a static model checker that unrolls
loops based on a predetermined (low) bound, whereas the loop in the
code needs to be executed a large number of times for the bug to be
triggered, BA is unable to build a formula for the failure at hand.
Unfortunately, increasing the number of times loops are unrolled is
not a viable solution, as it causes the number of encoded paths to
explode and results in the solver timing out.

Although this is just one bug in one program, and we cannot claim
generality of the results, we find the results very encouraging. They
provide evidence that our approach could make formula-based debugging
applicable to larger programs and real-world faults.

\begin{table}[t]
  \caption{Average time for processing \textsf{tcas} faults 
    using BA and OFC and two different solvers.}
\label{tab:solver}
\centering
\begin{scriptsize}
\begin{tabular}{|l|l|l|l|}
  \hline
  \scriptsize BA\_Yices & \scriptsize BA\_Z3 & \scriptsize OFC\_Yices & \scriptsize OFC\_Z3\\
  \hline
  187s & 166s & 48s & 26s \\
  \hline
\end{tabular}
\end{scriptsize}
\end{table}

\subsubsection{RQ4---Impact of Solver}

RQ4 aims to assess whether our results may depend on the use of a
specific solver. (For example, Yices may be optimized for the
wpMAX-SAT problem and give an unfair advantage to techniques that use
CW.) To answer RQ4, we focused on the results presented in
Table~\ref{tab:combined}, replaced Yices with another state-of-the-art
solver, Z3~\cite{z3}, and recomputed the results using this new
solver.  Because Z3 does not currently support wpMAX-SAT, we were only
able to use Z3 for the two techniques that do not use CW: BA and
OFC. Table~\ref{tab:solver} shows these results, side-by-side with the
earlier result we obtained using Yices.

From the results in the table, we can observe different values but a
similar trend for the results obtained using the two solvers. In both
cases, OFC can significantly improve the efficiency of formula-based
debugging (significance level of 0.05). Although this is just one
study involving two solvers, and thus we cannot claim general validity
of our results, it provides initial evidence that the improvements we
measured do not depend on the specific solver used.

\subsection{Threats to Validity}

In addition to the usual internal and external validity threats, a
specific one is that we implemented BugAssist, one of the techniques
against which we compare, ourselves. Unfortunately, we were not able to
use the command-line implementation of BugAssist from its authors, and
the Eclipse plugin was problematic to use in a programmatic way.
Moreover, the tool's source code was not available, which we needed to
integrate CW and BugAssist for our evaluation.

Another threat to validity is that we performed our evaluation on two
simple programs and a module of a real open source project. Therefore,
our results may not generalize. However, our main goal was to
investigate whether CW and OFC could improve the state of the art in
formula-based debugging, so using the same programs used in related
work allowed us to directly compare our techniques with such work.




\section{Related Work}
\label{sec:related-work}


Our work is closely related to formula-based debugging techniques. In
particular, OFC builds on BugAssist~\cite{Jose2011}, which encodes a
faulty program as an unsatisfiable Boolean formula, uses a MAX-SAT
solver to find maximal sets of satisfiable clauses in this formula,
and reports the complement sets of clauses as potential causes of the
error. The dual of MAX-SAT, that is the problem of computing minimal
unsatisfiable subsets (or unsatisfiable cores), can also be leveraged
in a similar way to identify potentially faulty statements, as done by
Torlak, Vaziri, and Dolby~\cite{Torlak2010}.  This kind of techniques
have the advantage of performing debugging in a principled way, but
tend to rely on exhaustive exploration of (a bounded version of) the
program state, which can dramatically limit their scalability.  OFC,
by operating on demand, can produce results that are at least as good
as those produced by these techniques at a fraction of the cost.
Moreover, by working on a single path at a time, OFC can directly
benefit from various dynamic optimizations. Finally, CW
leverages the additional information provided by passing test cases,
which are not considered by most existing techniques in this arena.

Another related approach, called Error Invariants, transforms program
entities on a single failing execution into a path
formula~\cite{Ermis2012}.  This technique leverages Craig interpolants
to find the points in the failing trace where the state is modified in
a way that affects the final outcome of the execution.  The statements
in these points are then reported as potential causes of the failure.
This technique cannot handle control-flow related faults because, as
also recognized by the authors, it does not encode control-flow
information in its formula.  To address this limitation, in followup
work the authors developed a version of their approach that encodes
partial control-flow information into the path
formula~\cite{Christ13}; with this extension, their approach can
identify conditional statements that may be the cause of a failure.
However, compared to OFC's approach of suitably encoding SSA's $\phi$
functions, their approach generates much more complex preconditions,
that is, conjunctions of all predicates that a statement is control
dependent on. Conversely, our algorithm only needs to encode the
predicate that the $\phi$ function is directly dependent on. In
addition, the two approaches handle potentially faulty conditional
statements very differently.  OFC considers additional paths induced
by a possible modification of the faulty conditionals, and can
therefore identify additional faulty statements along these paths.
Their technique simply reports the identified conditionals to
developers, who may miss important information and produce a partial,
if not incorrect, fix (see Section~\ref{sec:considerations}).

Our work is also related to statistical fault localization techniques
(\eg \cite{jones2002, Liblit2003, liblit05, Chen2002, Abreu2008,
  Santelices2009, Liu2005, Artzi2010, Zhang2009, Baah2011, Naish2011,
  Gopinath2012, Jin2013}).  Although efficient, these techniques often
produce long lists of program entities with no context information,
which can limit their usefulness~\cite{parnin11jul}. In CW, we use the
results of statistical fault localization as a starting point to
inform formula-based debugging and guide the analysis. As our results
show, these can result in considerably more accurate (and more
informative) fault-localization results.

Other approaches for identifying potentially faulty statements are
static and dynamic slicing~\cite{Agrawal1990,Weiser1981} and delta
debugging~\cite{Cleve2005, Zeller2002TSE}.
These approaches are orthogonal to ours and to formula-based debugging
techniques in general, and can be leveraged to achieve further
improvements.

Finally, automated repair techniques (\eg
\cite{dallmeier2009generating, legoues-tse2012, jeffrey2009bugfix})
are related to our work. However, these techniques are also mostly
orthogonal to fault-localization approaches, as they require some form
of fault localization as a starting point. (One exception is Angelic
Debugging, by Chandra and colleagues~\cite{chandra2011}, which
combines fault localization and a limited form of repair.)  In this
sense, we believe that the information produced by our approach could
be used to guide the automated repair generation performed by these
techniques, which is something that we plan to investigate in future
work.

\section{Conclusion and Future Work}
\label{sec:concl-future-work}

We presented clause weighting (CW) and on-demand formula computation
(OFC), two ways to improve existing formula-based debugging techniques
and mitigate some of their limitations. CW incorporates the
(previously ignored) information provided by passing test cases into
formula-based debugging techniques to improve their accuracy. OFC is a
novel formula-based debugging algorithm that, by operating on demand,
can analyze a small fraction of a faulty program and yet compute the
same results that would be computed analyzing the whole program, at a
much higher cost.

To evaluate CW and OFC, we performed an empirical study. In the study,
we assessed the improvements that CW and OFC can achieve over a
state-of-the-art formula-based debugging approach. Our results show
that CW and OFC can improve the accuracy, efficiency, and scalability
of the considered approach. Although our results are still
preliminary, we believe that they represent an important first step
towards further research in this important and promising area.

Our plans for future work are to perform additional studies, including
user studies, to further show the usefulness of our approach.
We will also explore the possibility of applying our on-demand
algorithm to other types of formula-based techniques, such as those
based on single-trace analysis (\eg \cite{Christ13, Ermis2012}), to
assess whether we can achieve similar, or even better improvements on
such techniques.
Finally, we will investigate whether formula-based debugging
techniques can help automated program repair.  Intuitively, the
clauses in the CoMSSs produced by the former should be able to inform
and guide the latter in finding or synthesizing suitable repairs.


\balance

\bibliographystyle{IEEEtran}
\bibliography{paper}

%
%

\end{document}